\documentclass[twocolumn,9pt]{article} 

\usepackage[square,numbers,sort&compress,comma]{natbib}

\usepackage{amsmath}
\usepackage{amssymb}
\usepackage{caption}
\usepackage{graphicx}
\usepackage{latexsym}
\usepackage{times}
\usepackage[pagewise]{lineno}
\usepackage{hyperref}
\usepackage{mhchem}[version=4]

\usepackage{cleveref}
\usepackage{microtype}
\usepackage{booktabs}
\usepackage{subcaption}
\usepackage{transparent}
\usepackage{xargs}
\usepackage{todonotes}
\newcommandx{\unsure}[2][1=]{\todo[linecolor=red,backgroundcolor=red!25,bordercolor=red,#1]{#2}}
\newcommandx{\change}[2][1=]{\todo[linecolor=blue,backgroundcolor=blue!25,bordercolor=blue,#1]{#2}}
\newcommandx{\info}[2][1=]{\todo[linecolor=green,backgroundcolor=green!25,bordercolor=green,#1]{#2}}
\newcommandx{\improvement}[2][1=]{\todo[linecolor=yellow,backgroundcolor=yellow!25,bordercolor=yellow,#1]{#2}}
\newcommandx{\thiswillnotshow}[2][1=]{\todo[disable,#1]{#2}}
\usepackage{comment}

\usepackage{xcolor}

\topmargin - 12pt 
\renewenvironment{abstract}%
              {
               \small
               {\bfseries \abstractname}
               \par
               \vspace{10pt}
              }

\renewcommand\abstractname{Abstract}

\newcommand{\nomenclature}
              [1]
              {
               \bgroup
               \flushleft
               \small\bf
               #1
               \par
               \egroup
              }

\renewcommand{\section}
              [1]
              {
               \bgroup
               \flushleft
               \small\bf
               \refstepcounter{section}
               \arabic{section}. #1
               \par
               \egroup
              }

\renewcommand{\subsection}
              [1]
              {
               \bgroup
               \flushleft
               \small\em
               \refstepcounter{subsection}
               \arabic{section}.
               \arabic{subsection}. #1
               \par
               \egroup
              }

\renewcommand{\subsubsection}
              [1]
              {
               \bgroup
               \flushleft
               \small\em
               \refstepcounter{subsubsection}
               \arabic{section}.
               \arabic{subsection}.
               \arabic{subsubsection}. #1
               \par
               \egroup
              }

  \newcommand{\acknowledgement}
              [1]
              {
               \bgroup
               \flushleft
               \small\bf
               #1
               \par
               \egroup
              }

  \newcommand{\sectionbib}
              [1]
              {
               \bgroup
               \flushleft
               \small\bf
               #1
               \par
               \egroup
              }

\newcommand{\ZePe}{\ensuremath{\mathit{Ze}/\mathit{Pe}}}

\begin{document}



\small
\baselineskip 10pt


\setcounter{page}{1}
\title{\LARGE \bf Scaling Laws for Thermodiffusively Unstable Lean
  Premixed Turbulent Hydrogen-Air Flames } \author{{\large
    M.~Gauding$^{a,*, \dagger}$, T.~Lehmann$^{a, \dagger}$,
    T.L.~Howarth$^{a,b}$, L.~Berger$^{a,c,d}$, M.~Rieth$^{e}$,
    A.~Gruber$^{f,g},$}\\ {\large W.~Song$^f$, J.H.~Chen$^{e}$,
    M.~Day$^g$, A.~Attili$^h$,
    E.F.~Hunt$^{i}$, A.J.~Aspden$^{i}$, H.~Pitsch$^{a}$}\\[10pt]
  {\footnotesize \em $^a$Institute for Combustion Technology, RWTH Aachen University, 52062 Aachen, Germany }\\[-5pt]
  {\footnotesize \em $^b$Department of Aeronautical and Automotive Engineering, Loughborough University, Loughborough LE11 3TU, UK}\\[-5pt]
  {\footnotesize \em $^c$Thermo and Fluid Dynamics (FLOW), Faculty of Engineering, Vrije Universiteit Brussel, Brussels 1050, Belgium}\\[-5pt]
  {\footnotesize \em $^d$Brussels Institute for Thermal-Fluid Systems and Clean Energy (BRITE), VUB-ULB, Brussels 1050, Belgium}\\[-5pt]
  {\footnotesize \em $^e$Combustion Research Facility, Sandia National Laboratories, Livermore, CA, USA}\\[-5pt]
  {\footnotesize \em $^f$Department of Energy and Process Engineering, Norwegian University of Science and Technology, Trondheim, Norway}\\[-5pt]
  {\footnotesize \em $^g$SINTEF Energy Research, Trondheim, Norway}\\[-5pt]
  {\footnotesize \em $^h$Institute for Multiscale Thermofluids, School of Engineering, University of Edinburgh, Edinburgh EH9 3FD, UK}\\[-5pt]
  {\footnotesize \em $^i$School of Engineering, Newcastle University, Newcastle-Upon-Tyne, Tyne and Wear NE1 7RU, UK}\\[-5pt]
}

\date{}  

\twocolumn[\begin{@twocolumnfalse}
\maketitle
\rule{\textwidth}{0.5pt}
\vspace{-5pt}

\begin{abstract} 
  Lean premixed hydrogen-air flames are strongly affected by
  thermodiffusive (TD) instabilities, which can alter the flame
  structure and enhance the local reactivity many-fold. 
 Two recent models (Howarth et al.~(Combust.~Flame 253, 2023) and Rieth et   al.~(Combust.~Flame 250, 2023)) describe the  scaling of the stretch factor using different measures of the TD response. The first uses the second-order coefficient of the linear dispersion relation, called $\omega_2$.  The second uses the ratio of the Zel’dovich to the Peclet number ($\mathit{Ze}/\mathit{Pe}$), relating   chemical reactivity and transport processes in the flame. 
  Using a comprehensive set of 91 direct
  numerical simulation (DNS) cases spanning a wide range of pressures,
  equivalence ratios, turbulence intensities, and flow configurations,
  both model formulations are systematically evaluated and an adapted formulation is proposed. The analysis of the governing non-dimensional groups reveals a scaling behavior characterized by two distinct regimes. In the first regime, typically relevant for burner and gas turbine conditions,  both models reduce to an identical form that depends solely on the Karlovitz number and the stretch factor of laminar flames, independent of $\omega_2$ or $\mathit{Ze}/\mathit{Pe}$. In the second regime, characterized by ultra-low flame speeds, the explicit consideration of $\omega_2$ or the ratio
  $\mathit{Ze}/\mathit{Pe}$ is required for accurate scaling. In both regimes, the two models predict the DNS data reasonably well and reduce to the same functional form of non-dimensional groups, indicating a consistency in their underlying physical description.
  
\end{abstract}

\vspace{5pt}

{\bf Novelty and significance statement}
The novelty of this work is the systematic evaluation and consolidation of two existing combustion models for turbulent lean hydrogen flames  using a comprehensive dataset of 91 direct numerical simulations covering a wide range of thermodynamic conditions and flame configurations. This study reconciles both formulations by demonstrating that both models are governed by the same non-dimensional groups. Furthermore, this study assesses how well combustion models, developed in canonical flame configurations in homogeneous, isotropic turbulence, translate to technically relevant jet flames. 
The study is significant because reliable turbulence–chemistry interaction models for predicting the burning rate of turbulent lean hydrogen flames affected by thermodiffusive instabilities are still lacking, particularly under high pressure conditions relevant to gas turbines and combustion engines. The results contribute  toward  developing predictive combustion models, thereby enabling the model-based design of sustainable combustion systems.
\vspace{10pt}

\parbox{1.0\textwidth}{\footnotesize {\em Keywords:}
  hydrogen; turbulence-chemistry interactions; thermodiffusive
  instabilities; scaling laws} \rule{\textwidth}{0.5pt} *Corresponding
author. $\dagger$ Joint first authors. These authors contributed equally. \vspace{5pt}
\end{@twocolumnfalse}]

\section{Introduction\label{sec:introduction}} \addvspace{10pt} Lean
premixed hydrogen-air flames are prone to thermodiffusive (TD)
instabilities, which originate from an imbalance between species and
thermal diffusive fluxes. The high diffusivity of hydrogen results in
effective Lewis numbers well below unity and promotes the
amplification of flame front perturbations leading to the formation of
cellular structures. TD instabilities have a leading order impact on
flame dynamics making the flame locally hotter, thinner, and faster
\cite{aspden2011turbulence,howarth2023thermodiffusively}.  Under these
conditions, the local flame speed strongly depends on strain and curvature of the flame front and can exceed the unstretched laminar burning
velocity by several times \cite{berger2022intrinsicB}. The increase of
the local reactivity by TD instabilities is quantified by the
non-dimensional stretch factor $I_0$, defined by
\begin{equation}
  \label{eq:I01}
  I_0 = \frac{s_c}{s_L \Psi} \,,
\end{equation}
where $s_c$ is the consumption speed, $s_L$ is the flame speed of a
one-dimensional unstretched flame, and $\Psi$ is the flame surface
area wrinkling.

TD instabilities interact synergistically with turbulence due to
increased curvature and enhanced stretch rate that intensify the TD
response \cite{day2009turbulence, aspden2019towards,
  berger2022synergistic, lapenna2024synergistic}.  The physical
mechanisms governing turbulent thermodiffusively unstable flames are
not yet fully understood and remain challenging to model
\cite{macart2018effects, berger2025combustion, chu2025extended,
  hok2025thickened}. In particular, a first-principles model for the
stretch factor $I_0$ that explicitly accounts for the TD response
depending on the local turbulence characteristics and the
thermodynamic conditions is  lacking.  To address this gap,
recent studies by Howarth~et~al.~\cite{howarth2023thermodiffusively} and
Rieth~et~al.~\cite{rieth2023effect, rieth2025} have proposed novel
empirical scaling relations for the stretch factor $I_0$ in
thermodiffusively unstable turbulent hydrogen flames under lean
conditions.   Howarth~et~al.~\cite{howarth2023thermodiffusively} introduced a model
(hereafter referred to as the $\omega_2$-model) that incorporates the  instability parameter $\omega_2$ derived from linear stability theory
\cite{matalon2003hydrodynamic,matalon2009flame,altantzis2012hydrodynamic}. In
contrast, Rieth~et~al.~\cite{rieth2023effect, rieth2025} proposed an
alternative model (hereafter referred to as the
$\mathit{Ze}/\mathit{Pe}$-model) that employs the ratio of the
Zel’dovich to the Peclet number to model the TD response.

Low-order combustion models for turbulent lean premixed hydrogen flames that
incorporate the scaling relation for the stretch factor $I_0$ by
Howarth~et~al.~\cite{howarth2023thermodiffusively} have recently been developed by
Chu~et~al.~\cite{chu2025extended} and Ramirez~et~al.~\cite{ramirez2025flame}. These models
evaluate the consumption speed $s_c$ from Eq.~\eqref{eq:I01} together with a flame wrinkling model \cite{Peters_2000} and
account for the increase in local reactivity induced by TD
instabilities through an empirical relation for $I_0$. The reliability
of such approaches depends on the availability of a precise, robust,
and physically consistent model for $I_0$.

However, the aforementioned models for $I_0$ have so far been evaluated only
for relatively simple flame configurations, and a direct comparison
between them using the same data is still missing.  The $\omega_2$-model was validated for
flames in forced homogeneous isotropic turbulence across a broad range
of thermodynamic conditions, including highly unstable flames at low
equivalence ratios and elevated pressures.  The
$\mathit{Ze}/\mathit{Pe}$-model was assessed for flames in turbulent
shear flows and homogeneous isotropic decaying turbulence,
where it demonstrated good predictive capability.

The present work pursues three main objectives. The first objective  is to establish  model hypotheses to generalize and align the model formulations. The second objective is to  evaluate and assess both models on a joint data base covering a broad spectrum of thermodynamic and turbulence conditions for multiple flame
configurations, including turbulent jet flames. These flames are
influenced by shear effects, inhomogeneities, and axial variations in turbulence. It remains unclear how combustion models developed in
statistically homogeneous, isotropic turbulence perform when applied
to these complex flame configurations.  
The third objective is to investigate the non-dimensional groups
of the scaling relations in order to identify the underlying physical principles. This analysis aims to assess the validity range of the
models and to determine conditions under which both formulations
converge.

The remainder of the paper presents a detailed comparison and analysis
of both model formulations. Section 2 introduces the model parameters
for laminar and turbulent flames. Section 3 presents the simulation
data used in this study.  Section 4 analyzes and evaluates the scaling
relations for the $\omega_2$-model and $\mathit{Ze}/\mathit{Pe}$-model and
evaluates their applicability to turbulent jet flames. It is shown that specific flame conditions exist under which both model formulations provide similar information. Based on this analysis, both models are consolidated and an adapted scaling relation is proposed.

\section{Modeling the TD response\label{sec:models}} \addvspace{10pt}
\subsection{Instability parameter $\omega_2$}\addvspace{10pt}
 Extensive research has been conducted to quantify the dynamics of
thermodiffusively unstable flames.  Canonical freely-propagating
laminar flames have been simulated to analyze the dependence of TD
instabilities on the reactant conditions \cite{berger2022intrinsicA,
  berger2022intrinsicB, howarth2023thermodiffusively, rieth2023effect,
  creta2020propagation}.  
  It has been shown that the TD response of a
lean premixed hydrogen flame can be characterized through classical
stability analysis \cite{matalon2003hydrodynamic,matalon2009flame}
using the parameter~$\omega_2$ from the dispersion relation
$\tilde \omega = \omega_{\rm DL} \tilde k + \omega_2 \tilde k^2$,
where $\omega_{\rm DL}$ is the Darrieus-Landau coefficient and
$\tilde \omega$ and $\tilde k$ are the the normalized growth rate and
wavenumber, respectively. The parameter $\omega_2$ is given by
\begin{equation}
  \label{eq:omega2}
  \omega_2 = - \left( B_1 + \mathit{Ze}(\mathit{Le}_{\rm eff}-1)B_2 + \mathit{Pr} B_3 \right) \,,
\end{equation}
where the coefficients $B_i$ depend on the density ratio~$\sigma$ and
the temperature variation of the thermal conductivity $\lambda$. The
Zel'dovich number is denoted by $\mathit{Ze}$, $\mathit{Pr}$ is the
Prandtl number, and $\mathit{Le}_{\rm eff}$ is the effective Lewis
number evaluated as~\cite{altantzis2012hydrodynamic} 
\begin{equation}
    \mathit{Le}_{\rm eff} = 1 + \frac{\left(\mathit{Le}_{\mathrm{O_2}}-1\right) + \left(\mathit{Le}_{\mathrm{H_2}}-1\right)\mathcal{A}}{1+\mathcal{A}}
\end{equation}
with species' Lewis numbers calculated in the burnt mixture and $\mathcal{A}=1 + \mathit{Ze} \left(\phi^{-1}-1\right)$. Following Law and Sung~\cite{law2000}, the Zeldovich
number is defined as
\begin{equation}
  \mathit{Ze}=\frac{E}{R}\frac{T_{\rm b}-T_{\rm u}}{T_{\rm b}^2}\,,
	\label{eq:ZeldovichBase}
\end{equation}
where $T_u$ and $T_b$ refer to the temperature in the unburnt and
burnt mixture, respectively, $R$ is the universal gas constant and $E$
is the activation energy, given by
\begin{equation}
  \frac{E}{R}=-2\frac{\mathrm{d}\left(\ln\left(\rho_{\rm u} s_{\rm L}\right)\right)}{\mathrm{d}(1/T_{\rm b})}\,,
  \label{eq:ActivationEnergy}
\end{equation}
with $\rho_u$ being the density of the unburnt mixture. In this work,
the derivative in Eq.~(\ref{eq:ActivationEnergy}) is numerically evaluated by varying the mass fraction of \ce{N2}. It should be noted that both $\mathit{Le}_{\rm eff}$ and $\mathit{Ze}$ are sensitive to the evaluation method~\cite{lehmann2025comprehensive} and hence need to be used consistently in the model formulation and evaluation.  If the parameter $\omega_2$ is
negative, initial perturbations are stabilized through the TD response. In contrast, a positive value of $\omega_2$ signifies an intrinsically  unstable flame based on a combination of thermodiffusive and hydrodynamic effects.

In laminar flames, the TD response depends on the mixture and the thermodynamic conditions.
Howarth and Aspden~\cite{howarth2022empirical} demonstrated that for each combination of
unburnt temperature $T_u$ and equivalence ratio~$\phi$, there exists a
critical pressure $\Pi_c$, at which the TD instability reaches its
maximum.  The critical pressure~$\Pi_c$ is defined by
\begin{equation}
  \Pi_c = \left( \frac{20 \phi}{7-2 \Theta} \right)^{150/(21+10 \Theta)} \,,
\end{equation}
with $\Theta = T_u/300 {\rm K}$ and $\Pi_c = p/1 {\rm atm}$.
Howarth and Aspden~\cite{howarth2022empirical} observed that  scaling relations of flame speed, flame thickness, and local reactivity change across the critical pressure~$\Pi_c$. This is a direct consequence of a shift in reaction pathways with increasing pressure \cite{law2006propagation}.  With increasing pressure, the chain terminating termolecular reaction $\mathrm{H + O_2 + M \rightarrow HO_2 + M}$ becomes increasingly important, compared to the chain branching reaction $\mathrm{H + O_2 \rightarrow O + OH}$ thereby shifting the balance toward less reactivity. This is reflected by the increase in $\mathit{Ze}$. However, with further increasing pressures, the \ce{HO2} sub chemistry tends to produce more \ce{H2O2} which subsequently forms \ce{2 OH} via another termolecular reaction, increasing global reactivity again, and hence reduces $\mathit{Ze}$ and thermodiffusive instabilities~\cite{law2006propagation}.
Consequently, separate empirical scaling relations for the stretch factor under laminar conditions $I_0^*$ were identified for the regime below and above the critical pressure $\Pi_c$, hereafter referred to as the low- and the high-pressure
regimes:
\begin{equation}
  \label{eq:I0lam}
  I_0^* =
  \begin{cases}
    \exp(0.08 \omega_2) & \text{if}\; p \le \Pi_c,   \\
    1+0.47 \omega_2, & \text{otherwise} \,.
  \end{cases}
\end{equation}
It is shown in Section \ref{sec:I0} that the separation between the low- and high-pressure regimes persists in turbulent flames.
In this work, properties of three-dimensional, freely propagating laminar flames that may be affected by TD instability are denoted by an
asterisk. Note that these properties differ from values of laminar, unstretched one-dimensional flames. For thermodiffusively unstable flames, $\omega_2 > 0$ and
$I_0^* > 1$.

Note that conditions in the high-pressure regime are typically characterized by ultra-low flame speeds resulting from  a combination of ultra-lean equivalence ratio, low unburnt
temperature, and high pressure. For instance, typical gas turbine
operating conditions ($T_u = 700$~K, $p =20-30$~bar,
$\phi = 0.4$) are located in the low-pressure regime, whereas
conditions encountered in internal combustion engines with exhaust gas
recirculation, which also exacerbates instability, can extend into the high-pressure regime during
the combustion cycle \cite{Golc2026}.

  \subsection{ Peclet number $\mathit{Pe}$}\addvspace{10pt}
Besides $\omega_2$, other parameters have been employed to describe
the TD response.  Recently, Rieth~et~al.~\cite{rieth2025} proposed a novel
scaling for lean premixed hydrogen flames that relates the stretch
factor $I_0$ to the ratio $\mathit{Ze}/\mathit{Pe}$. Here,
 the Zel'dovich number $\mathit{Ze}$ characterizes 
the flame's reactivity, with larger $\mathit{Ze}$ corresponding to lower reactivity. The Peclet number~$\mathit{Pe}$ is defined as
the ratio of convective to molecular transport and can be understood as
a measure for the flame's propensity to become thermodiffusively
unstable. The Peclet number is evaluated from one-dimensional
unstretched flames as
\begin{equation}
  \label{eq:Pe}
  \mathit{Pe} = \frac{
    \left|\rho u \frac{d Y_{\rm H_2}}{d x} \right|_{\rm 1D,max}
  }{
    \left| \frac{d }{d x} \left( \rho \frac{W_{\rm H_2}}{W} D_{\rm H_2} \frac{d X_{\rm H_2}}{d x}  \right)  \right|_{\rm 1D,max}
  } \,,
\end{equation}
where $\rho$ is the density, $u$ is the velocity, $W_{\rm H_2}$ and
$D_{\rm H_2}$ are the molecular weight and diffusion coefficient of
$\rm H_2$, respectively, and $W$ is the mean molecular weight of the
mixture.  Rieth~et~al.~\cite{rieth2023effect} demonstrated that the ratio
$\mathit{Ze}/\mathit{Pe}$ accurately quantifies the flame's propensity
to develop TD instabilities. From direct numerical simulation (DNS)
data, they showed that the stretch factor
$I_0^*$ of two-dimensional, laminar, freely propagating flames scales almost linearly
with $\mathit{Ze}/\mathit{Pe}$ over a pressure range from 1 to 20~bar. This scaling relation is revisited in Section~\ref{sec:ZePe} and extended to three-dimensional flames and different pressure regimes.

  \subsection{Turbulence-chemistry interaction}\addvspace{10pt}
The parameters $I_0^*, $ $\omega_2$, and $\mathit{Ze}/\mathit{Pe}$
describe the thermophysical state of the flame and do not contain any
information about turbulence-chemistry interaction. Therefore, these
parameters are combined with the turbulent Karlovitz number to predict
the stretch factor in turbulent flames. The Karlovitz number
characterizes small-scale turbulence-chemistry interaction and is
defined as the ratio of the flame time $\tau_F$ to the Kolmogorov time
scale $\tau_\eta$. The Karlovitz
number can be defined as
\begin{equation}
  \label{eq:Ka}
  \mathit{Ka} = \sqrt{{\tilde \varepsilon}/{\varepsilon_F}},
\end{equation}
where $\varepsilon_F = s_L^3/l_F$ and $\tilde \varepsilon$ is the
Favre-averaged turbulent kinetic energy dissipation rate, given by
\begin{equation}
  \label{eq:edr}
  \tilde \varepsilon = \widetilde{\tau_{ij}'' \frac{\partial u_i''}{\partial x_j}} \,,
\end{equation}
where $\tau_{ij}$ is the viscous stress tensor.
The Karlovitz number is
evaluated in the unburnt mixture at the $C_{\rm H_2}=0.1$ iso-surface,
where $C_{\rm H_2}= 1-Y_{\rm H_2}/Y_{\rm H_2,u}$ is the progress variable based on the fuel mass fraction.

A refined definition was proposed by
\cite{aspden2011characterization}, in which the Karlovitz number
$\mathit{Ka}^*$ is defined using the burning velocity $s_L^*$ and the
flame thickness $l_F^*$ of the corresponding freely propagating
laminar flame, i.e.,
\begin{equation}
  \label{eq:Ka_star}
  \mathit{Ka}^* = \sqrt{\tilde \varepsilon/\varepsilon_F^*},
\end{equation}
with $\varepsilon_F^*= {s_L^*}^3/l_F^*$.  Empirical models for $s_L^*$ and $l_F^*$ have been developed by Howarth and Aspden~\cite{howarth2022empirical}.  

In this work, properties of laminar flames, i.e., $I_0^*$, $s_L^*$, and $l_F^*$, are extracted from simulation data of three-dimensional, laminar, freely-propagating flames, so that no additional modeling of these parameters is required. This approach allows the analysis to focus on the turbulence-chemistry interaction model.

\section{Simulation data\label{sec:data}}\addvspace{10pt}

\begin{table}
\centering
\caption{Parameters of the cases F1, F2, TSL, and D.  }
\label{tab:FIAB}
\setlength{\tabcolsep}{4pt}
\footnotesize
\begin{tabular}{lcccc}
\toprule
Case & F1 & F2 & TSL & D \\
\midrule
  $p$ (atm)      & 1 -- 40   & 1 -- 40   & 1--10   & 16 \\
  $T_u$ (K)      & 300 -- 700 & 300 -- 700 & 750     & 693 \\
  $\phi$         & 0.2 -- 0.57 & 0.2 -- 0.4 & 0.3     & 0.45 \\
  $\omega_2$     & 0 -- 27 & 2.8 -- 27 & 0 -- 4.5   & 3.15 \\
  $\mathit{Ka}$ &  1.3 -- 2580 & 1.7 -- 302 & 88 -- 189 & 4 -- 30  \\
  $\mathit{Ka}^*$& 1 -- 36   & 1 -- 36   & 43 -- 176 & 2.4 -- 92 \\
  $I_0$ & 1.3 -- 13.3 & 2.0 -- 13.3 & 2.0 -- 5.8 & 1.6 -- 2.4 \\
\bottomrule
\end{tabular}
\vspace{1eM}
\caption{Parameters of the jet flames.  }
\label{tab:cases}
\footnotesize
\setlength{\tabcolsep}{5pt}
\begin{tabular}{lccc}
  \toprule
  Case & 1atm & 5atm & 10atm \\
  \midrule
  $p$ (atm) & 1  & 5 & 10 \\
  $H$ (mm) & 8 & 2 & 1.5 \\
  $U$ (m/s) & 24& 19.2 & 10.6 \\
  $\omega_2$ & 5.94 & 19.9 & 29.8 \\
  $s_L^*$ (cm/s) & 32.3 & 25.4 &  16.5 \\
  $l_F^*$ ($\mu$m) & 471 & 112 & 102 \\
  $\mathit{Ka}$ & 13 -- 162 & 115 -- 1463 & 226 -- 1986 \\ 
  $\mathit{Ka^*}$ & 4.6 -- 57.6 & 3.8 -- 48.9 & 3.2 -- 69.5 \\
  $I_0$ & 1.6 -- 4.0   & 3.6 -- 12.8  & 6.1 -- 28.2 \\
\bottomrule
\end{tabular}
\end{table}

DNS data covering a wide range of thermodynamic and turbulence
conditions are used to assess the proposed scaling relations. The
dataset includes a total of 91 cases, comprising flames in forced (F1~\cite{howarth2023thermodiffusively}, F2~\cite{hunt2025thermodiffusively}) and decaying (D~\cite{rieth2025})
homogeneous isotropic turbulence, temporally evolving shear layers
(TSL~\cite{rieth2025}), and piloted jet flames in a slot configuration (Jet~\cite{howarth2026}),
with Karlovitz numbers $\mathit{Ka}$ ranging from 1.3 to 2580
($\mathit{Ka}^*$ between 1 and 176), pressures from 1 to 40~atm,
equivalence ratios from 0.2 to 0.57, and unburnt temperatures from 300
to 750~K. For the temporally evolving cases, statistics are sampled at
the instant when the peak of $I_0$ occurs. Parameters of the simulation
are given in Table~\ref{tab:FIAB}. A detailed list of all data points
is provided in the supplementary material.

Based on the definition of $\mathit{Ka}^*$, which accounts for
the thinning of thermodiffusively unstable flames, all conditions
considered in this work are located in the thin reaction zone regime.
The only exception is the TSL flame with $\omega_2 \approx 0$, which extends
into the distributed burning regime with $\mathit{Ka}^* = 176$.
In addition, also its $\mathit{Pe}$ and $\mathit{Ze}$ numbers differ
substantially from those of the other cases, resulting in a somewhat different scaling behavior of this case. These quantities are also
provided in the list in the supplementary material.

Table~\ref{tab:cases} summarizes the simulation parameters of the jet
flames.  The jet Reynolds number is fixed at 10,000, while the pressure is
increased from 1~to~10~atm. The unburnt mixture has a temperature of
$T_u=298$~K and an equivalence ratio of $\phi=0.4$. The slot width~$H$
corresponds to 17 characteristic flame thicknesses
$l_F^* = l_F \exp(-0.06\omega_2)$, where $l_F$ is the unstretched
laminar flame thickness \cite{howarth2022empirical}. The bulk velocity~$U$ is adjusted to maintain
a constant Reynolds number for all cases.  Statistics are averaged temporally
and spatially along the spanwise direction and presented along the
streamwise coordinate.  Further details on the configuration are
provided by Howarth et al.~\cite{howarth2026}.

For the jet flames, Table~\ref{tab:cases} shows that statistics vary over wide ranges in the flame. Specifically, both the Karlovitz
number and the stretch factor decay along the streamwise direction. For the configuration considered, increasing the pressure leads
to higher values of both~$I_0$ and~$\mathit{Ka}$. Before reaching the flame tip region, the streamwise
decrease of the stretch factor results from the decay of the Karlovitz
number.

\begin{figure}
	\centering
	\includegraphics[width=\linewidth]{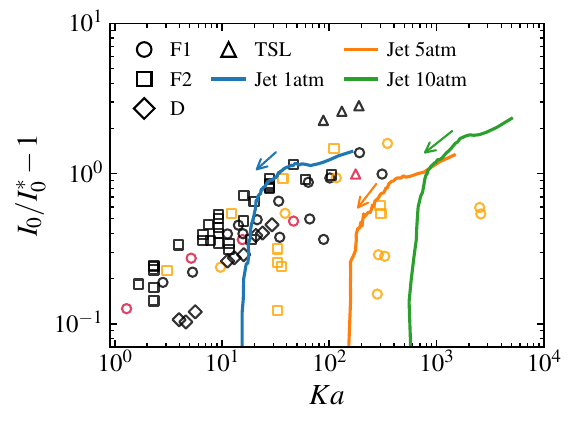}
	\caption{ Enhancement of reactivity $I_0$ from the laminar
          freely propagating flame as a function of the Karlovitz
          number for all cases. Conditions in the high-pressure regime ($p> \Pi_c$) are highlighted in orange. Cases with $\omega_2 \approx 0$ (thermodiffusively stable conditions) are marked in red. The streamwise direction of the jet flames is indicated by an arrow. }
	\label{fig:data}
\end{figure}

Figure \ref{fig:data} shows the dependence of $I_0/I_0^* - 1$ on the
Karlovitz number $\mathit{Ka}$ for all cases. This ratio represents the relative
enhancement of the stretch factor by turbulence compared to laminar
freely propagating flames. For the dataset considered, considerable
scatter among the data points is evident: $\mathit{Ka}$ spans
more than three orders of magnitude, while $I_0/I_0^* - 1$ reaches
values as large as three. This results in local flame speeds that are
up to 10 times faster compared to TD stable flames. Notably, even the
TD-stable flames with $\omega_2 \approx 0 $ and $I_0^*\approx 1$ exhibit a substantial increase in
local flame speed as indicated by $I_0/I_0^* - 1 > 0$. The physical origin of this behavior and implications for modeling are  discussed in Sections~\ref{sec:omega2} and \ref{sec:low-p}.

\section{Scaling relations for the stretch factor $I_0$\label{sec:I0}}\addvspace{10pt}
In the following, models for $I_0$ are systematically evaluated using
the DNS data outlined in Section~\ref{sec:data}. As models for $I_0$ derived from first principles are not available, empirical scaling relations are required to  capture the relevant physical effects. The overarching modeling hypotheses are that $I_0$  approaches $I_0^*$ in the limit of vanishing turbulence, i.e., $\mathit{Ka}^* \to 0$, and that the additional increase in $I_0$ due to interactions of the flame with turbulence are captured through the Karlovitz number $\mathit{Ka}^*$. The Karlovitz number is defined using the burning velocity $s_L^*$ and flame thickness $l_F^*$ of thermodiffusively unstable, freely propagating laminar flames to account for modifications of the flame structure due to TD instability. This definition of the Karlovitz number was first proposed by Aspden et al. \cite{aspden2011characterization}. Further,  $\mathit{Ka}^*$ is modulated by a prefactor that depends on $\omega_2$ (Howarth et al.~\cite{howarth2023thermodiffusively})  or $\mathit{Ze}/\mathit{Pe}$ (Rieth et al.~\cite{rieth2025}) to account for the strength of the TD instability. Note that the original model proposed by Rieth et al.~\cite{rieth2025} is reformulated in this work to align with these modeling hypotheses. 

 Chu~et~al.~\cite{chu2025extended} implemented an $\omega_2$-based closure in a Reynolds-averaged Navier-Stokes (RANS) framework, where the scaling relation can be applied directly using modeled turbulence quantities. Using a local definition of the Karlovitz number  makes the model compatible with large eddy simulations (LES), as demonstrated by Ramirez~et~al.~\cite{ramirez2025flame}. 
combustion closures.
 However, the predictive use of the model requires careful validation for the specific operating conditions and flame configuration. The present study supports such developments by assessing the accuracy and limitations of the required scaling relations for $I_0$.

The model parameters are fitted against the DNS data described in Section \ref{sec:data}.  The jet flame data are not used for calibration of the model parameters,
but rather provide an out-of-sample assessment of whether  a model developed from flame-in-a-box simulations generalizes to technically relevant flame configurations that include shear, inhomogeneity,  and strong  variations of turbulence.

\subsection{Scaling of the $\omega_2$-model\label{sec:omega2}}\addvspace{10pt}
Howarth et al.~\cite{howarth2023thermodiffusively} proposed an empirical scaling
relation for the stretch factor $I_0$ that accounts for TD
instabilities, expressed as
\begin{equation}
  \label{eq:TLH}
  I_0 = \left( 1 + q_\mathrm{H} \exp(n_\mathrm{H} \omega_2) {\mathit{Ka}^*}^{m_\mathrm{H}} \right) I_0^* \,,
\end{equation}
where $q_\mathrm{H}$, $n_\mathrm{H}$, and $m_\mathrm{H}$ are fitting parameters.
Equation~\eqref{eq:TLH} considers TD instabilities through the
parameter $\omega_2$ and the factor $I_0^*$, which is the stretch
factor of freely propagating laminar flames.  This formulation is
consistent with the laminar limit as $I_0 \to I_0^*$ when
$\mathit{Ka}^* \to 0$. In the original study by
Howarth et al.~\cite{howarth2023thermodiffusively}, the model was fitted against
dataset F1, with fitting exponent $m_\mathrm{H}$ fixed at~0.5.

Figure~\ref{fig:scaling_tlh} shows a least-squares fit of
Eq.~\eqref{eq:TLH} to the complete dataset, excluding the jet
flames. The quality of the fit is quantified by the mean absolute percentage error (MAPE), which measures the average percentage difference between the predicted and true values. The jet flames are excluded from the fitting procedure to evaluate the
model's predictive capability for technically relevant
configurations. Interestingly, a $\mathit{Ka}$-exponent close to 0.5 is observed. The negative
$\omega_2$ factor in the exponent in Eq.~\eqref{eq:TLH}, see Fig.~\ref{fig:scaling_tlh}, reflects that turbulence has
a reduced influence on $I_0$ when the TD instability is already
strong \cite{howarth2023thermodiffusively}. 

A generally good collapse of the data points is
observed. However, larger deviations from the scaling appear for the
thermodiffusively stable flames (marked in red), the TSL flames (back triangles), and the jet flames
at elevated pressure (orange and green lines).  The jet flames exhibit substantial variations in $I_0$ and
$\mathit{Ka}$ along the axial direction (see~Fig.~\ref{fig:data}) that are not predicted
correctly by the model. However, the strong decay of $I_0$ occurs in the flame-tip region, where the scaling relations are not expected to hold.

Conditions with $\omega_2 \approx 0$ (highlighted in red in the scaling plots) correspond to thermodiffusively stable flames. Under these conditions,
$I_0$ can still exceed unity, indicating a thermodiffusive response
with locally enhanced reactivity. This effect originates from
turbulence-induced flame curvature rather than intrinsic instability
and is therefore not dependent on~$\omega_2$. Specifically, turbulence
generates curved flame elements, which induce differential and
preferential diffusion when the Lewis number deviates from unity. For
hydrogen flames with $\mathit{Le} \approx 0.3$, this mechanism leads to
locally enriched and hotter regions in convex flame elements, resulting in faster burning and an overall increase
of $I_0$ above unity. This case is not fully covered by
the model fit as shown in Fig.~\ref{fig:scaling_tlh}.

\subsection{Scaling of the $\mathit{Ze}/\mathit{Pe}$-model\label{sec:ZePe}}\addvspace{10pt}

\begin{figure}
  \centering
  \includegraphics[width=\linewidth]{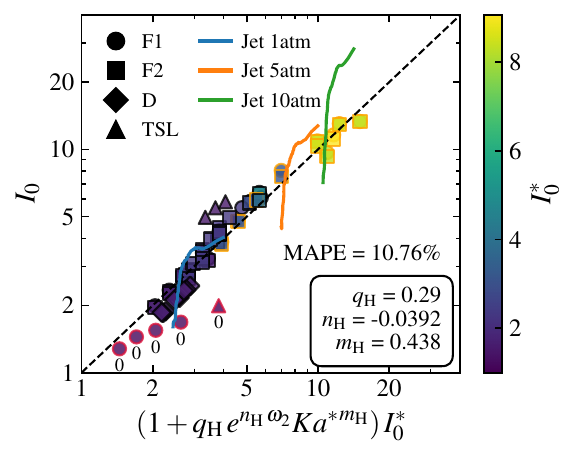}
  \caption{Evaluation of the scaling relation given by
    Eq.~\eqref{eq:TLH}.  Data with $\omega_2 \approx 0$ are highlighted in
    red, and data belonging to the high-pressure regime ($p > \Pi_c$)
    are highlighted in yellow. MAPE is the mean-absolute percentage
    error.}
  \label{fig:scaling_tlh}
  \includegraphics[width=\linewidth]{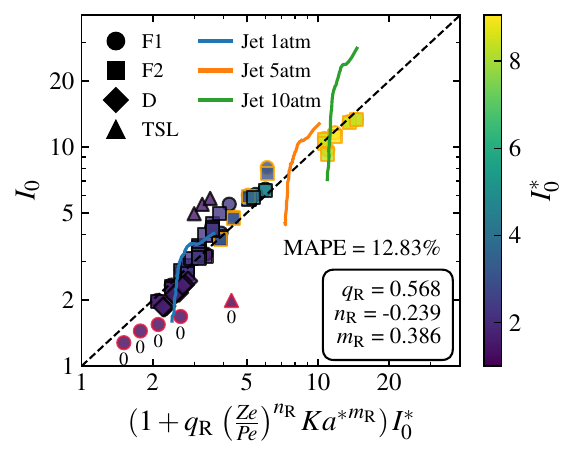}
  \caption{Evaluation of the scaling relation from
    Eq.~\eqref{eq:MRmod}. Data with $\omega_2 \approx 0$ are highlighted in
    red, and data belonging to the high-pressure regime ($p > \Pi_c$)
    are highlighted in yellow. MAPE is the mean-absolute percentage
    error. }
  \label{fig:scaling_mr}
\end{figure}

Rieth et al.~\cite{rieth2025} proposed an alternative scaling relation for the
stretch factor, given by
\begin{equation}
  \label{eq:MR}
  I_0 = 1 + q_\mathrm{R} \left( \frac{\mathit{Ze}}{\mathit{Pe}} \right)^{n_\mathrm{R}} {\mathit{Ka}}^{m_\mathrm{R}} \,,
\end{equation}
which accounts for TD instabilities through the parameter
$\mathit{Ze}/\mathit{Pe}$. $q_\mathrm{R}$, $n_\mathrm{R}$, and $m_\mathrm{R}$ are model specific fitting parameters.  Equation~\eqref{eq:MR} does not satisfy the
laminar limit, where $I_0$ should equal $I_0^*$. Therefore, the original model formulation given by
Eq.~\eqref{eq:MR} is adapted in this study analogous to the $\omega_2$-model in
the form:
\begin{equation}
  \label{eq:MRmod}
  I_0 = \left[ 1 + q_\mathrm{R} \left( \frac{\mathit{Ze}}{\mathit{Pe}} \right)^{n_\mathrm{R}} {\mathit{Ka}^*}^{m_\mathrm{R}} \right] I_0^* \,,
\end{equation}
which substitutes the exponential term $\exp(n_\mathrm{H} \omega_2)$ from Eq.~\eqref{eq:TLH} with
$( \mathit{Ze}/\mathit{Pe})^{n_\mathrm{R}}$, and ensures that $I_0 \to I_0^*$ as
$\mathit{Ka}^* \to 0$. The difference between Eq.~\eqref{eq:MR} and
Eq.~\eqref{eq:MRmod} becomes significant when $I_0^* \gg 1$. Furthermore, Eq.~\eqref{eq:MRmod} is formulated in terms of $\mathit{Ka}^*$ to account for the local increase in flame speed and the resulting thinning of the flame in thermodiffusively unstable flames \cite{aspden2011turbulence}. Equation~\eqref{eq:MRmod} is referred to as
modified \ZePe-model. A least-squares fit of Eq.~\eqref{eq:MRmod} to the complete dataset,
excluding the jet flames, is shown in Fig.~\ref{fig:scaling_mr}. The resulting
scaling exhibits an accuracy comparable to that of the
$\omega_2$-model shown in Fig.~\ref{fig:scaling_tlh}.

\begin{figure}
  \centering
  \includegraphics[width=\linewidth]{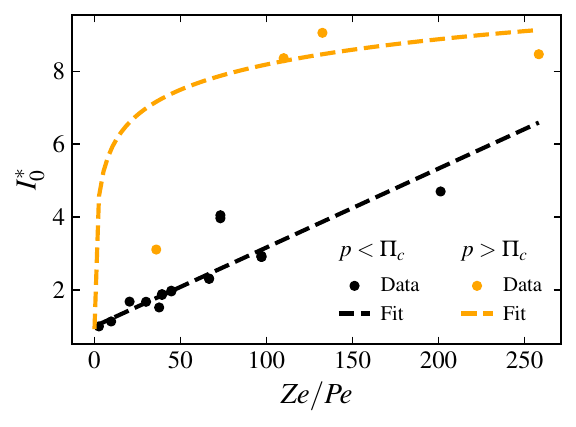}
  \caption{Parameterization of $I_0^*$ with $\mathit{Ze}/\mathit{Pe}$. Conditions in the low-pressure regime are shown in black, conditions in the high-pressure regime are indicated in orange. }
  \label{fig:I0_ZePe}
\end{figure}

\begin{figure*}[h!]
  \centering
  \begin{subfigure}{0.48\textwidth}
    \includegraphics[width=\linewidth]{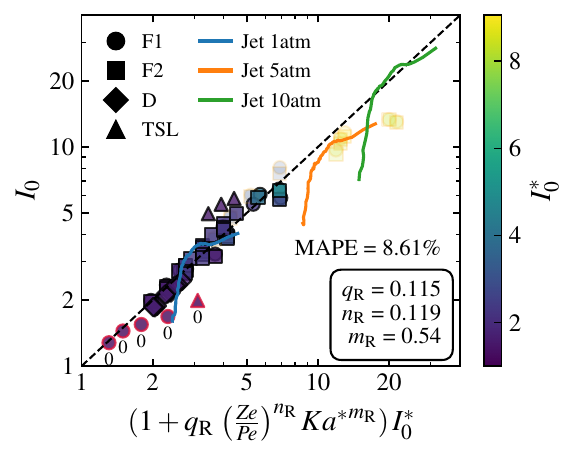}
    \caption{}
    \label{fig:MR_LP}
  \end{subfigure}
  \hfill
  \begin{subfigure}{0.48\textwidth}
    \includegraphics[width=\linewidth]{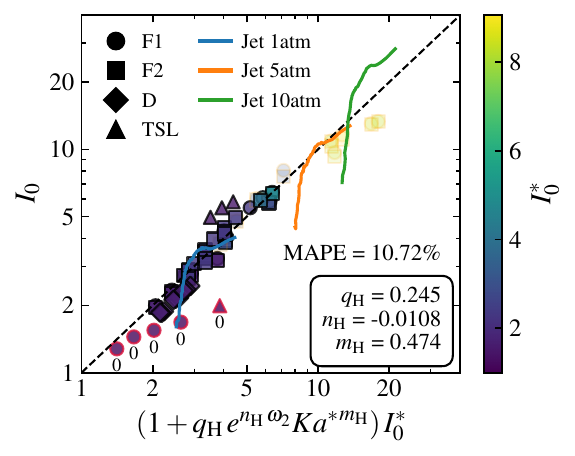}
    \caption{}
    \label{fig:TLH_LP}
  \end{subfigure}
  \begin{subfigure}{0.48\textwidth}
    \includegraphics[width=\linewidth]{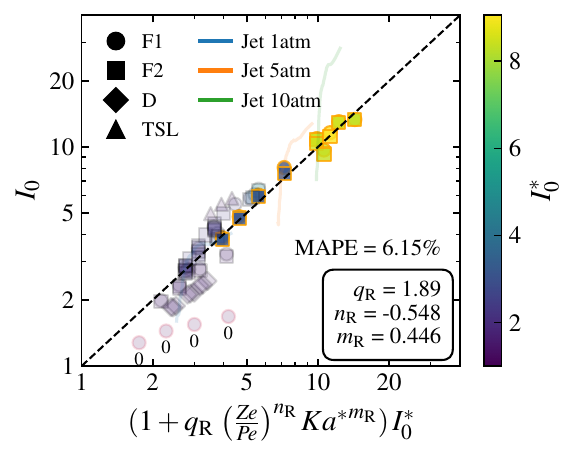}
    \caption{}
    \label{fig:MR_HP}
  \end{subfigure}
  \hfill
  \begin{subfigure}{0.48\textwidth}
    \includegraphics[width=\linewidth]{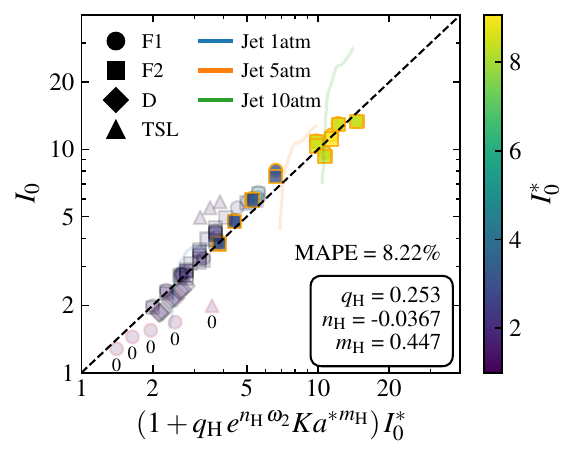}
    \caption{}
    \label{fig:TLH_HP}
  \end{subfigure}
  \caption{Evaluation of the scaling relations in the low- (top row) and high-pressure (bottom row) regimes.  Left column: modified \ZePe-model from Eq.~\eqref{eq:MRmod}. 
  Right column: $\omega_2$-model from Eq.~\eqref{eq:TLH}. 
  Coloring as in Fig.~\ref{fig:scaling_tlh}.}
\end{figure*}

For a self-consistent and complete model formulation, an expression for $I_0^*$ in terms of $\mathit{Ze}$ and $\mathit{Pe}$ would have to be provided.
 A corresponding model was proposed by Rieth et al.~\cite{rieth2023effect} using simulations of two-dimensional, laminar, freely-propagating flames. In the present study, the model by Rieth~et~al. is extended to the high-pressure regime and reformulated using  data from three-dimensional flames considered here. Figure~\ref{fig:I0_ZePe} reveals that a parameterization of $I_0^*$ in terms of $\mathit{Ze}/\mathit{Pe}$ follows two distinct scaling relations in the low- and high-pressure regimes. This result extends the study by Rieth et al.~\cite{rieth2023effect}  to different pressure regimes and  is consistent with the parameterization of $I_0^*$ with $\omega_2$, see Eq.~\eqref{eq:I0lam}. A least-squares fit yields:
\begin{equation}
  \label{eq:I0lam_R}
  I_0^* =
  \begin{cases}
    1 + 0.022  \left(\mathit{Ze}/\mathit{Pe}\right) & \text{if}\; p \le \Pi_c,   \\
    \ln\left(36.1 \left(\mathit{Ze}/\mathit{Pe}\right)\right)  & \text{otherwise} \,.
  \end{cases}
\end{equation}
Comparing  Eq.~\eqref{eq:I0lam_R} with Eq.~\eqref{eq:I0lam} suggests an exponential relationship between the parameters used in the two models, which explains the different prefactors in the $\mathit{Ka}^*$-term of the scaling laws.  This observation represents an important step toward reconciling the two scaling relations.  However, to ensure consistency and comparability between the scaling relations, the laminar flame properties used in this work are extracted directly from the simulation data rather than being evaluated using Eq.~\eqref{eq:I0lam_R}.

\section{Consolidation of scaling laws}\addvspace{10pt}

The findings from Sec.~\ref{sec:I0} indicate that representing the entire dataset
with a single set of non-dimensional groups, i.e., $\omega_2$ or $\mathit{Ze}$ and $\mathit{Pe}$, is challenging. The existence of separate scaling relations for the low- and high-pressure regimes (see Eqs.~\eqref{eq:I0lam} and \eqref{eq:I0lam_R}) therefore motivates to treat these regimes separately.  
This split of pressure regimes is performed in the following and is shown to improve the predictive capability of both models.

\subsection{Scaling laws for the low-pressure regime\label{sec:low-p}} \addvspace{10pt}

Figure~\ref{fig:MR_LP} shows the modified
$\mathit{Ze}/\mathit{Pe}$-model fitted to the data in the low-pressure regime,
demonstrating a significantly improved collapse across all conditions, reaching a MAPE of 8.61\% with parameters $q_\mathrm{R} = 0.115$, $n_\mathrm{R} = 0.119$, and $m_\mathrm{R} = 0.540$. A
similar result is obtained for the $\omega_2$-model, which yields a MAPE of 10.72\% with with parameters $q_\mathrm{H} = 0.245$, $n_\mathrm{H} = -0.0108$, and $m_\mathrm{H} = 0.474$ (see
Fig.~\ref{fig:TLH_LP}). It is interesting to note that both exponents $m_\mathrm{R}$ and $m_\mathrm{H}$ are very
close to 0.5 for both models, as suggested by Howarth et al.~\cite{howarth2023thermodiffusively}.
Considering the prefactors $(\mathit{Ze}/\mathit{Pe})^{n_\mathrm{R}}$ and
$\exp(n_\mathrm{H}\omega_2)$, it is found that, in the low-pressure regime, both of
these vary by less than 20\% across all cases due to the small
magnitude of the parameters $n_\mathrm{R}$ and $n_\mathrm{H}$. Under these conditions,
the prefactors have a small sensitivity on the resulting scaling relation and can be regarded as approximately constant, implying
that the $\omega_2$-model and the modified
$\mathit{Ze}/\mathit{Pe}$-model yield effectively the same scaling
behavior. Consequently, both models reduce to the same functional form, given by
\begin{equation}
  I_0 = \left( 1 + q {\mathit{Ka}^*}^m \right) I_0^* \,,
\end{equation}
highlighting their consistency in the technically most relevant low-pressure regime. The evaluation of this scaling relation yields  similar performance,  as  shown in the supplementary material. This is a significant result, as it reduces the number of parameters required for modeling and thereby simplifies the
implementation of the scaling relation in combustion models. At the
same time, the model still accounts for local variations in turbulence in the flame
through the Karlovitz number.

The data from the jet flames are captured reasonably well by
the $\mathit{Ze}/\mathit{Pe}$-model, indicating that the model has
some predictive capability to complex 
flames. The $\omega_2$-model shows a comparable accuracy for the jet
flames at 1~and 5~atm but performs less well for the~10~atm case. 
For both models, the accuracy decreases toward the flame tip, which may be explained by  flame-flame interactions and expansion effects that induce strong fluctuations in the local Karlovitz number not accounted for in the model.

Compared to the $\omega_2$-model, the modified $\mathit{Ze}/\mathit{Pe}$-model provides a higher predictive accuracy for the thermodiffusively stable conditions with $\omega_2 \approx 0$, which contributes to the slightly lower MAPE. 
This difference can be traced to the modulation terms in the two formulations. As $\omega_2 \rightarrow 0$, the factor $\exp(n_H\omega_2)$ approaches unity and therefore no longer carries flame-specific information at the stability boundary. In contrast, $(\mathit{Ze}/\mathit{Pe})^{n_R}$ remains non-unity and can still discriminate between near-stable flame conditions through differences in reactivity and transport.
The relatively large deviation of the TSL case with $\omega_2 \approx 0$ can be attributed to its very large Karlovitz number, $\mathit{Ka}^*=176$, which is the only case in this study that is  located towards the distributed burning regime.

\subsection{Scaling laws for the high-pressure regime}\addvspace{10pt}

Figure~\ref{fig:MR_HP} shows the scaling of the modified
$\mathit{Ze}/\mathit{Pe}$-model in the high-pressure regime. A good
collapse is observed for cases within the high-pressure regime (MAPE of 6.15\%).
The fitted parameters are $q_\mathrm{R} = 1.89$, $n_\mathrm{R} = -0.548$, and
$m_\mathrm{R} = 0.446$, indicating that, in contrast to the low-pressure regime,
the prefactor $(\mathit{Ze}/\mathit{Pe})^{n_R}$ cannot be considered
constant. This
result highlights the existence of two distinct sets of
non-dimensional groups valid in either the low- or high-pressure
regime. Similar behavior is obtained for the $\omega_2$-model
with parameters $q_\mathrm{H} = 0.253$, $n_\mathrm{H} = -0.0367$, and
$m_\mathrm{H} = 0.447$ and a MAPE of 8.22\% (see Fig.~\ref{fig:TLH_HP}). Still, in both models the $\mathit{Ka}^*$ exponent is close to 0.5.

\begin{figure}
	\centering
	\includegraphics[width=\linewidth]{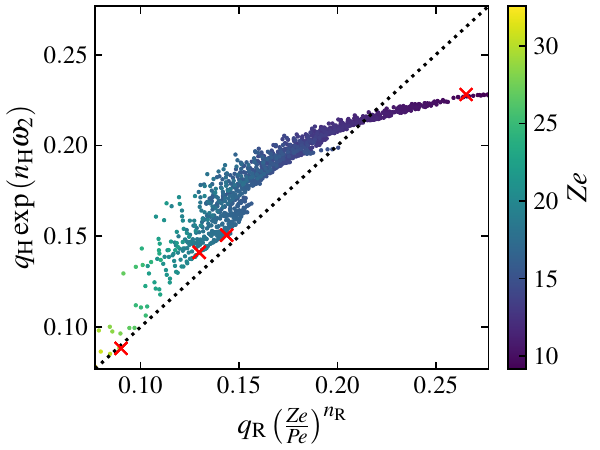}
	\caption{Comparison of the scaling coefficients in the high-pressure regime ($p>\Pi_c$) with one-dimensional  simulations of unstretched premixed flames over a wide parameter space. Data points  are colored by $\mathit{Ze}$. Red crosses indicate the thermodynamic conditions of the high-pressure branch cases used for fitting of the scaling relations. Details of these conditions can be found in the supplementary material.}
	\label{fig:ModelComp}
\end{figure}

\begin{figure}
	\centering
	\includegraphics[width=\linewidth]{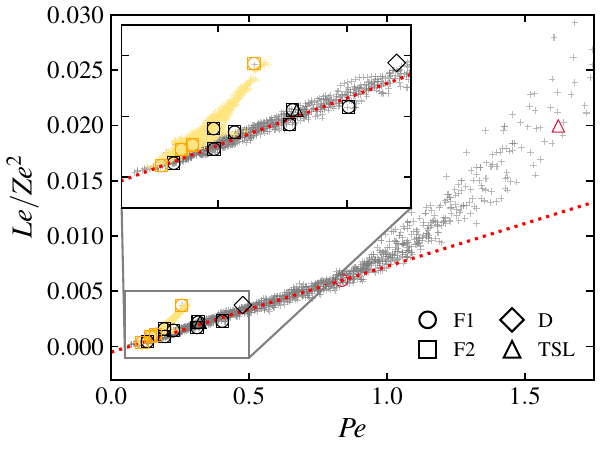}
	\caption{Scaling relation between $\mathit{Le}/\mathit{Ze}^2$ and $\mathit{Pe}$, where gray and yellow markers denote one-dimensional unstretched flame
          simulations over a wide parameter space. The turbulent flames considered in this study are indicated by open symbols. Black and yellow markers correspond to the low- and high-pressure regimes, respectively. Most of these conditions are located within the highlighted rectangle, where $\mathit{Ze}$ is large. Red symbols indicate thermodiffusively stable conditions with $\omega_2 \approx 0$.  The outlier from the scaling relation (red triangle) is the TSL flames with $\omega_2 \approx 0$.}
	\label{fig:PeLe}
\end{figure}

To demonstrate the similarity of the models within the high-pressure regime, 
the scaling coefficients are compared. The preceding analysis showed that the model parameters $m_\mathrm{R}=0.446$ and $m_\mathrm{H}=0.447$ are nearly identical in both models, which allows the scaling coefficients of the \ZePe- and $\omega_2$ models to be equated as:
\begin{equation}
	\label{eq:model_sim}
 	q_{\rm R} \left(\frac{\mathit{Ze}}{\mathit{Pe}}\right)^{n_{\rm
            R}} \approx q_{\rm H} \exp\left(n_{\rm
          H}\omega_2\right)\,.
\end{equation}
 Figure~\ref{fig:ModelComp} compares  the left- and right-hand sides of Eq.~\eqref{eq:model_sim}, evaluated on one-dimensional unstretched flame solutions over a wide range of thermo-chemical conditions. The red crosses indicate the thermodynamic conditions of the cases from the high-pressure branch of the DNS data set used for fitting. For large  $\mathit{Ze}$, a consistency in the trend between the two model coefficients can be observed. At small $\mathit{Ze}$, this correlation becomes invalid. In the next step, analytical evidence for this correlation is provided  by rewriting the definition of $\mathit{Pe}$ in
Eq.~\eqref{eq:Pe} using the mass-based diffusion
flux $\rho D \mathrm{d}Y/\mathrm{d}x$ and
assuming that the characteristic length scales of convection and
diffusion correspond to  the thickness of the diffusive layer and the reaction zone
$l_{D}$ and $l_{R}$, respectively. This yields
\begin{equation}
  \label{eq:PeLe_scaling}
	\mathit{Pe}\propto \frac{\rho_u s_L \frac{1}{l_{D}}}{\rho D
          \left(\frac{1}{l_{R}}\right)^2} \propto \frac{\lambda /
          c_p}{\rho D}\left(\frac{l_{R}}{l_{D}}\right)^2 \propto
        \frac{\mathit{Le}}{\mathit{Ze}^{2}}\,,
\end{equation}
where the relations
$\rho_u s_L \propto \lambda / (c_p l_D)$~\cite{bechtold1987, law2005}
and $l_D/l_R \propto \mathit{Ze}$~\cite{law2000} are
applied, under the assumption of large $\mathit{Ze}$. Figure~\ref{fig:PeLe} illustrates $\mathit{Le}/\mathit{Ze}^2$ as a function of $\mathit{Pe}$ for one-dimensional unstretched premixed flames over a wide parameter space, supporting the validity of Eq.~\eqref{eq:PeLe_scaling}. For large $\mathit{Ze}$, i.e., small $1/\mathit{Ze}^2$, $\mathit{Le}/\mathit{Ze}^2$ scales linearly with $\mathit{Pe}$. Furthermore, Fig.~\ref{fig:PeLe} reveals two distinct branches for the low- and high-pressure regimes.
The scaling relation for the high-pressure
regime ($p>\Pi_c$), represented by the yellow branch in
Fig.~\ref{fig:PeLe}, requires a different proportionality constant
compared to the low-pressure regime ($p<\Pi_c$).  This observation not only
demonstrates the existence of two distinct pressure regimes but also
explains the reduced predictive accuracy of the modified \ZePe-model for conditions with $p > \Pi_c$ observed in 
Fig.~\ref{fig:scaling_mr}. Similar as in Fig.~\ref{fig:ModelComp}, this correlation becomes invalid at low $\mathit{Ze}$, i.e., high $1/\mathit{Ze}^2$, which is grounded in the assumption of large $\mathit{Ze}$ in the derivation of Eq.~(\ref{eq:PeLe_scaling}). Most conditions within the scope of this work satisfy the condition of large $\mathit{Ze}$.

Substituting Eq.~\eqref{eq:PeLe_scaling} and the fitted model
parameters into Eq.~\eqref{eq:model_sim} and rearranging yields
\begin{equation}
	\label{eq:model_sim_2}
	C \mathit{Ze}\mathit{Le}^{-1/3} \approx \exp\left(0.022\omega_2\right)\,,
\end{equation}
where $C$ absorbs the model constants $q_\mathrm{R}$ and $q_\mathrm{H}$ and the
proportionality factor of Eq.~\eqref{eq:PeLe_scaling}. The complete derivation of Eq.~\eqref{eq:model_sim_2} is provided in Sec.~3 of the supplementary material. By applying
first-order Taylor series expansions to the left- and right-hand sides, Eq.~\eqref{eq:model_sim_2} can be
expressed as
\begin{equation}
\begin{aligned}
	&C\,\mathit{Ze}\left(C^\dagger_1 - C^\dagger_2\,\mathit{Le}\right)
	\approx 1 + {}\\
	&\quad 0.022\left[-B_1 + B_2\,\mathit{Ze}(1-\mathit{Le}) + B_3\,\mathit{Pr}\right]\,,
\end{aligned}
\label{eq:model_sim_final}
\end{equation}
where $C^\dagger_i$ are constants from the Taylor series expansion.  Note that
$\mathit{Le}$ does not vary significantly for the cases considered, since $\phi \le 0.56$,
and that the argument of the exponential function is generally small,
due to the small value of $n_{\rm H}$, which justifies the applied expansions.
Furthermore, an expansion around $\mathit{Le}=0.25$ gives
$C^\dagger_1 = C^\dagger_2$, such that it can be factorized. 
The resulting relation suggests that both models depend on similar combinations of non-dimensional groups and exhibit the same leading-order dependence on $\mathit{Ze}(1-\mathit{Le})$.
This should not be interpreted as a mathematical identity or a  general conversion formula between 
$\omega_2$ and $\mathit{Ze}/\mathit{Pe}$. Rather, the purpose is to show that both parameterizations involve related non-dimensional groups in the high-pressure regime.  Therefore, for this analysis, second-order terms in the Taylor expansion can be neglected. An extended discussion further justifying this assumption is provided in the supplementary material.

\section{Conclusion\label{sec:conclusion}} \addvspace{10pt} 

Two recently proposed scaling models (the $\omega_2$- and \ZePe-models)
for predicting the stretch factor $I_0$ in thermodiffusively unstable
lean premixed hydrogen–air flames were systematically evaluated using
data from 91 DNS cases covering a wide range of thermodynamic
conditions and flame configurations. To this end,  the \ZePe-model was reformulated to satisfy the physical limit of laminar flames by introducing~$I_0^*$, and to account for the enhanced flame speed of thermodiffusive unstable flames by replacing~$\mathit{Ka}$ by $\mathit{Ka}^*$.
With these modifications, both models capture the primary
trends of the DNS data, although clear deviations for certain
conditions occur. It is also shown that both  models have some predictive capability for the stretch factor in turbulent jet flames, although the accuracy decreases toward the flame tip. 

The analysis indicated that representing the entire dataset with a
single set of non-dimensional groups is difficult due to the existence
of two distinct pressure regimes. This observation suggests that the distinction into two pressure regimes extends from laminar to turbulent flames.  Consequently, the analysis was
conducted separately for the low- and high-pressure regimes.  In the
low-pressure regime, both models converge and take an analogous form
that depends solely on $\mathit{Ka}^*$ and $I_0^*$.  In contrast, in
the high-pressure regime, the parameters $\omega_2$ or
$\mathit{Ze}/\mathit{Pe}$ are essential beyond their role in determining $I_0^*$ to achieve an adequate
collapse of the data. However, dimensional analysis showed that both model formulations are effectively governed by the same functional relation of the non-dimensional groups $\mathit{Ze}(1-\mathit{Le})$. In other words,  although both models are formulated in terms of different parameters, this study demonstrates  that these parameters are correlated and can be expressed by the same non-dimensional groups, leading to comparable predictive accuracy.

\acknowledgement{CRediT authorship contribution statement} \addvspace{10pt}

\textbf{MG}: Writing – original draft, methodology, analysis, conceptualization, project administration. \textbf{TL}: Writing – original draft, methodology, analysis, conceptualization, data curation. \textbf{TLH, LB, MR, AG, JHC, AJA}: Writing – review \& editing, methodology,  analysis, conceptualization, data curation. \textbf{WS, MD, EFH}: Writing – review \& editing, data curation. \textbf{AA}: Writing – review \& editing, methodology, analysis, conceptualization. \textbf{HP}: Writing – review \& editing, methodology, analysis, conceptualization, project administration.

\acknowledgement{Declaration of competing interest} \addvspace{10pt}
The authors declare that they have no known competing 
interests.

 \acknowledgement{Acknowledgments} \addvspace{10pt} 
 %
%
%
%

RWTH Aachen University: German Research Foundation (DFG) within the Priority Program SPP 2419 (523874889), European Research Council Advanced Grant (HYDROGENATE, 101054894), IRTG 2983 HyPotential (516338899); Gauss Center for Supercomputing e.V., Jülich Supercomputing Center, and Leibniz Supercomputing Center. Sandia: US Department of Energy (DOE), Office of Basic Energy Sciences. Sandia is a multimission laboratory managed and operated by National Technology and Engineering Solutions of Sandia, LLC. for the DOE NNSA  under contract DE-NA-0003525. Newcastle University: EPSRC (EP/W034506/1). Norway: European Union’s Horizon Europe research and innovation program  (HyPowerGT, 01136656),  UNINETT Sigma2 (BETZY supercomputer: nn8035k, ns8035k).
\footnotesize
\baselineskip 9pt

\clearpage
\thispagestyle{empty}
\bibliographystyle{proci}
\bibliography{PCI_LaTeX}


\newpage

\small
\baselineskip 10pt


\end{document}